\documentclass[reprint,amsfonts, amssymb, amsmath,  showkeys, nofootinbib,pra, superscriptaddress, twocolumn,longbibliography]{revtex4-2}
\usepackage[normalem]{ulem}
\usepackage{float}
\makeatletter
\let\newfloat\newfloat@ltx
\makeatother

\usepackage[english]{babel}

\usepackage[utf8]{inputenc}
\usepackage{graphics}
\usepackage{selinput}
\usepackage{comment}

\usepackage{comment}
\usepackage{braket}
\usepackage{amsthm}
\usepackage{mathtools}
\usepackage{physics}
\usepackage{xcolor}
\usepackage{graphicx}
\usepackage[left=16mm,right=16mm,top=35mm,columnsep=15pt]{geometry} 
\usepackage{adjustbox}
\usepackage{placeins}
\usepackage[T1]{fontenc}
\usepackage{lipsum}
\usepackage{csquotes}

\usepackage[linesnumbered,ruled,vlined]{algorithm2e}
\SetKwInput{kwInit}{Init}

\def\ad{^{\dagger}}

\usepackage[makeroom]{cancel}
\usepackage[toc,page]{appendix}
\usepackage[colorlinks=true,citecolor=blue,linkcolor=magenta]{hyperref}
\usepackage[capitalise]{cleveref}

\newcommand{\poly}{\operatorname{poly}}

\newcommand{\OC}{\mathcal{O}}

\renewcommand{\geq}{\geqslant}
\renewcommand{\leq}{\leqslant}
\newcommand{\mte}[2]{\langle#1|#2|#1\rangle }

\renewcommand{\vec}[1]{\boldsymbol{#1}}  

\newcommand{\bs}{\textsf{BS}}

\newcommand{\thv}{\vec{\theta}}

\newcommand{\losalamos}{Theoretical Division, Los Alamos National Laboratory, Los Alamos, New Mexico 87545, USA}

\def\be{\begin{equation}}
\def\ee{\end{equation}}
\def\bs{\begin{split}}
\def\e{\end{split}}
\def\ba{\begin{eqnarray}}
\def\bea{\begin{eqnarray}}

\def\tea{\end{eqnarray}}
\def\ea{\end{eqnarray}}
\def\eea{\end{eqnarray}}

\newtheorem{definition}{Definition}

\usepackage{amssymb}
\usepackage{dsfont}

\def\be{\begin{equation}}
\def\te{\end{equation}}
\def\ee{\end{equation}}
\def\ba{\begin{eqnarray}}
\def\bea{\begin{eqnarray}}

\def\tea{\end{eqnarray}}
\def\ea{\end{eqnarray}}
\def\eea{\end{eqnarray}}

\newcommand{\dimg}{\text{dim}(\mathfrak{g})}
\newcommand{\ham}{H}

\newcommand{\permuteop}[2]{#2 #1 #2^{-1}}
\newcommand{\permutestate}[2]{#2 \ket{#1}}
\newcommand{\aut}[1][\ham]{\text{Aut}(#1)}
\newcommand{\autinv}{\mathcal{E}(\aut)}

\newcommand{\orbit}{\mathcal{O}}
\newcommand{\orbitcircuit}{ORB circuit}
\newcommand{\eorbit}{\mathcal{O}^{e}}
\newcommand{\graph}{\mathcal{G}}

\newcommand{\old}[1]{}
\newcommand{\swap}{\text{SWAP}}
\newcommand\mbb[1]{\mathbb{#1}}

\begin{document}

\title{Building spatial symmetries into parameterized quantum circuits for faster training}

\author{Fr\'{e}d\'{e}ric Sauvage}
\thanks{The two first authors contributed equally.}
\affiliation{\losalamos}

\author{Mart\'{i}n Larocca}
\thanks{The two first authors contributed equally.}
\affiliation{\losalamos}
\affiliation{Center for Nonlinear Studies, Los Alamos National Laboratory, Los Alamos, New Mexico 87545, USA}

\author{Patrick J. Coles}
\affiliation{\losalamos}
\affiliation{Quantum Science Center, Oak Ridge, TN 37931, USA}

\author{M. Cerezo}
\email{cerezo@lanl.gov} 
\affiliation{Information Sciences, Los Alamos National Laboratory, Los Alamos, NM 87545, USA}
\affiliation{Quantum Science Center, Oak Ridge, TN 37931, USA}

\begin{abstract}
Practical success of quantum learning models hinges on having a suitable structure for the parameterized quantum circuit. Such structure is defined both by the types of gates employed and by the correlations of their parameters. While much research has been devoted to devising adequate gate-sets, typically respecting some symmetries of the problem, very little is known about how their parameters should be structured. In this work, we show that an ideal parameter structure naturally emerges when carefully considering spatial symmetries (i.e., the symmetries that are permutations of parts of the system under study). Namely, we consider the automorphism group of the problem Hamiltonian, leading us to develop a circuit construction that is equivariant under this symmetry group. The benefits of our novel circuit structure, called ORB, are numerically probed in several ground-state problems. We find a consistent improvement (in terms of circuit depth, number of parameters required, and gradient magnitudes) compared to literature circuit constructions. 
\end{abstract}

\maketitle

\section{Introduction}

With the advent of quantum technologies, opportunities to explore learning models blending quantum and classical components have emerged. Research in this direction was initiated by the development of the Variational Quantum Eigensolver (VQE)~\cite{peruzzo2014variational} for electronic and nuclear structure and the Quantum Approximate Optimization Algorithm (QAOA)~\cite{farhi2014quantum} for combinatorial optimization. Both algorithms frame practical problems as an optimization over quantum states prepared by a Parameterized Quantum Circuit (PQC). These proposals were quickly followed by many more, for applications ranging from linear systems to factoring to dynamical simulation, forming the field of Variational Quantum Algorithms (VQAs)~\cite{cerezo2020variationalreview,bharti2021noisy}.

In VQAs, the practical success of a given algorithm hinges on the choice of a suitable structure for the PQC employed (commonly referred as the ansatz). To date, two main generic families of circuits have emerged.
On one hand, the Hardware-Efficient Ansatzes (HEAs)~\cite{kandala2017hardware} propose to make the most of a given quantum platform by parameterizing each individual gates natively supported by the underlying device. 
These circuits offer great flexibity in that they can express a wide range of potential solutions in shallow depths. 
However, such enhanced expressibility comes at the expense of trainability issues limiting the extent to which these circuits can be optimized at scale~\cite{mcclean2018barren,cerezo2021cost,holmes2021connecting,larocca2021diagnosing}.

On the other hand, problem--inspired ansatzes are designed to respect the structure of the problem they aim to solve and inherently restrain the space of solutions explored. This includes the Hamiltonian Variational Ansatz (HVA)~\cite{wecker2015progress,ho2019efficient,wiersema2020exploring,cade2020strategies} and the quantum alternating operator ansatz~\cite{farhi2014quantum,hadfield2019quantum}. (We will refer to them collectively as HVA-like circuits as they share the same structure.) While successful in mitigating some trainability issues~\cite{wiersema2020exploring,larocca2021diagnosing}, these constructions rely on correlating parameter values over many gates, yielding relatively deep circuits and suggesting that they could be realized in shallower depths. We note that such deep circuits lead to noise-induced trainability issues~\cite{wang2020noise}.

Striking the right balance between flexible circuit structures that could approximate the solution of a problem in shallow depths, but, spanning only a restricted (but still relevant) search space remains a great challenge. Finding this ``Goldilocks'' construction for an ansatz, which has sufficiently many independent parameters per layer to be realized in shallow depth, but still is problem-inspired enough to maintain a low expressibility, is the topic of our work.

Following conceptual advances in the field of classical machine learning~\cite{cohen2016group,kondor2018generalization,bronstein2021geometric}, emphasis has recently been put forward to develop the framework of geometric quantum machine learning, where one structures 
quantum learning models based on symmetries of the task to be solved~\cite{larocca2022group,meyer2022exploiting,skolik2022equivariant}. In the context of quantum circuit design, respecting (all or some of) the symmetries of a problem can be achieved by an appropriate choice of the type of gates constituting the circuit employed, and also, imposing correlation patterns in the parameters of such gates.
Symmetry-compliant gate-sets have received considerable attention and are already well understood for groups of symmetries commonly encountered in the realm of quantum physics~\cite{liu2019variational,seki2020symmetry,gard2020efficient,zheng2021speeding} (see also \cite{marvian2022restrictions} for limitations entailed by gate-locality constraints). In contrast, principled ways to arrange gate parameters are still lacking.

\begin{figure}
	\includegraphics[width=0.48\textwidth]{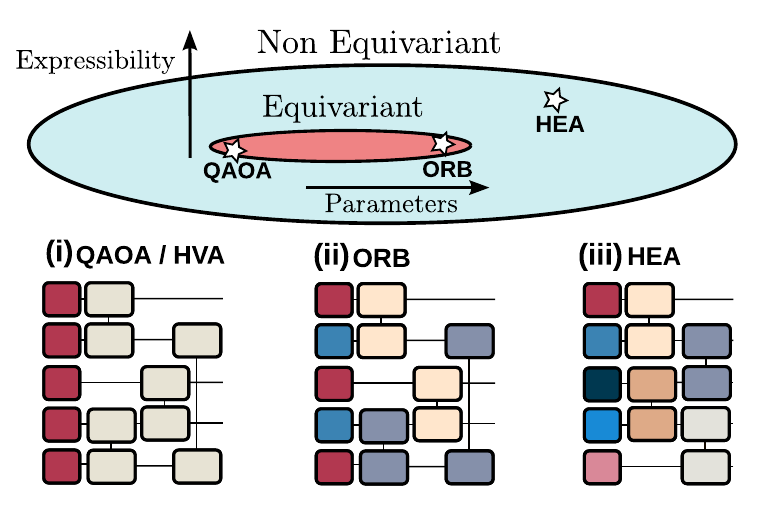}
	\caption{\textbf{Comparison of \orbitcircuit\   with literature ansatzes.}
    In this work, we propose a quantum circuit construction (dubbed \orbitcircuit ) and demonstrate its potential in a variety of ground-state preparation tasks. The parameter structure of the circuits proposed is taylored to respect the spatial symmetries of the problem to be solved while maximizing the number of independent parameters.
    When compared to problem--inspired ansatzes (i), such as the HVA and QAOA, \orbitcircuit s (ii) can pack many more free parameters per layer with the same (or marginally increased) expressibility, and thus, benefit from the same trainability guarantees but in shorter depths. 
    When compared to problem--agnostic circuits like the HEA (iii), \orbitcircuit s (ii) exhibit much more restrained expresssibility, so as to span only the relevant (Equivariant) solution space. Circuit layouts for these different constructions are sketched in the bottom panels. Identical (distinct) colors correspond to identical (independent) parameters. This accounts for 2, 4 and 9 independent (sometimes also termed as free or uncorrelated) parameters for the circuits (i), (ii) and (iii) respectively}.    
	\label{fig:overall}
\end{figure}

In this work, we explore the design of quantum circuits with a particular emphasis on the choice of an appropriate parameter structure. An optimal choice is found to naturally emerge when carefully considering the spatial symmetries of the system under study. This leads us to develop a general circuit construction that is dubbed \orbitcircuit\ (or ORB ansatz)  as it correlates gate parameters belonging to a same \emph{orbit} according to the symmetries of the problem. As illustrated in \Cref{fig:overall}, this construction aims at bridging the gap between (and combining the respective strengths of) HEA--like circuits and HVA--like ones.

For a variety of ground state preparation tasks, we numerically verify that the methodology proposed allows to maximize the number of free parameters per circuit layer without impeding the trainability of the circuits. In particular, we demonstrate that: 
(i) When compared to HVA--like constructions, \orbitcircuit\ can pack many more free parameters per layer (up to a quadratic factor in the problem size), while ensuring a similar expressibility, resulting in significantly shorter \orbitcircuit s. (Here, expressibility is measured in terms of the dimension of the circuit's Lie algebra~\cite{larocca2021diagnosing}.)
(ii) When compared to ansatzes with independent parameters, the circuits presented achieve similar performances with fewer parameters and overall better trainability at scale.
(iii) Somewhat surprisingly, and in contrast with current practices of QAOA~\cite{farhi2014quantum}, this suggests that independent parameters should be adopted in many problems of graph optimizations and beyond.

The rest of this work is structured as follows. 
After introducing the necessary background in~\cref{sec:tech}, we motivate and detail the construction of \orbitcircuit s in \cref{sec:orb}. These are then applied and benchmarked in~\cref{sec:res} over four different models including studies of the transverse-field Ising model in $1$D and $2$D, problems of Max-Cut, and the $J_1$--$J_2$ Heisenberg model on a square lattice.

\section{Preliminaries}\label{sec:tech}
We start this preliminary section by formalizing further the problem at hand in~\cref{sec:pb}. 
Then, we detail the main principles guiding the design of quantum circuits with a focus on trainability, including discussions on the expressibility of quantum circuits in~\cref{sec:expr} and on the limits imposed by noise in~\cref{sec:noise}.

\subsection{Problem at hand}\label{sec:pb}
In this work, we consider a generic learning task ubiquitous to the field of VQAs: given a Hamiltonian $\ham$, one is interested in preparing its ground state.  In the context of VQAs, this task is formulated as an optimization over states $\ket{\psi(\thv)}$ that are realized by applying a parameterized unitary $U(\thv)$ -- implemented by means of a quantum circuit -- to a fixed initial state $\ket{\psi_0}$. 
The circuit parameters $\thv$ are then optimized to minimize the energy of the realized state, that is, the expectation value
\begin{equation}\label{eq:exp}
    E(\thv) = \mte{\psi(\thv)}{\ham} = \mte{\psi_{0}}{U(\thv)\ad \ham U(\thv)}\,.
\end{equation}

\subsubsection{Circuit structure}
At the core of the success of any task of ground state preparation lies the structure of the circuit employed.
Quantum circuits are typically structured as a composition of repeated layers  
\begin{equation}
    U(\thv) = \prod^L_{l=1} V(\thv^{(l)})\,
\end{equation}
where $V$ denotes a parametrized unitary corresponding to the action of a layer, with parameters $\thv^{(l)}$ varying from one layer (indexed as $l \in [1,L]$) to another.  
These layers can be further decomposed in terms of gates, typically acting on single qubits or two qubits, that are either fixed gates or parameterized ones of the form $R_G(\theta)=\exp[-i \theta G]$ for a generator $G$ (a traceless Hermitan operator) and a variable parameter $\theta$.
We denote as $\mathcal{G}=\set{G_g}$ the set of all the generators involved in a circuit.

\subsubsection{HEA circuits}
In the most general case, the parameters associated to each of the parameterized gates involved in a given layer can be taken to be uncorrelated. For instance, the HEA ansatz employed in~\cite{kandala2017hardware} has layers of the form
\begin{equation}\label{eq:HEA}
    V_{\text{HEA}}(\thv^{(l)}) = U_{ent} \times \prod^n_{i=1} R_{Z}(\thv_{i,1}^{l}) R_{X}(\thv_{i,2}^{l}) R_{Z}(\thv_{i,3}^{l})
\end{equation}
composed of a fixed entangling unitary $U_{ent}$ and parametrized single-qubit gates taken to be either Pauli $X_i$ or Pauli $Z_i$ rotations acting on the qubits $i=1, \hdots, n$. 
Notably, each of these $3n$ rotations per layer is controlled by an individual parameter $\thv_{i,o\in \set{1,2,3}}^{(l)}$ that can be tuned independently of any other rotation.
In such a case, neither the choice of the gates nor the choice of the parameter structure depends on the underlying problem to be solved.

\subsubsection{HVA-like circuits}
As an alternative paradigm, in the circuit layers of the HVA-like anstazes (encompassing HVA circuits~\cite{wecker2015progress,ho2019efficient,wiersema2020exploring}, QAOA circuits~\cite{farhi2014quantum} and their generalization~\cite{hadfield2019quantum}) most of the parameters are correlated.
That is, a single parameter value is used for many of the gates appearing in the same layer. 
For instance, a layer of the the alternating ansatz in Refs.~\cite{wecker2015progress,ho2019efficient,wiersema2020exploring} takes the form of 
\begin{equation}\label{eq:alt}
    V_{\text{HVA}}(\thv^{(l)}) = \prod^n_{i=1} R_{X_i}(\thv_{1}^{(l)}) \prod_{(i,j) \in \mathcal{I}} R_{Z_i Z_j}(\thv_{2}^{(l)}),
\end{equation}
where each of the $n$ qubits are rotated along the $X$--axis with the same angle $\thv_{1}^{(l)}$, 
and where each pairs of qubits $(i,j)$ belonging to some set $\mathcal{I}$, defined in correspondence to an underlying Hamiltonian, are subjected to $ZZ$ interactions with the same strength $\thv_{2}^{(l)}$. Going beyond these two extreme cases of independent and (almost fully) correlated parameters is the topic of this work.

\old{
\begin{equation}
    V_{\text{HVA}}(\thv^l) = \prod^n_{i=1} \text{exp}[-i \thv_{1}^{l} X_i] \prod_{\langle i,j \rangle} \text{exp}[-i \thv_{2}^{l} Z_iZ_j],
\end{equation}
}

\subsection{Trainability of parameterized quantum circuits}\label{sec:train}

Following the many proposals of VQAs ansatzes, it has been realized that their application to large problem sizes would, in some cases, quickly result in trainability issues rendering the optimization of the circuit parameters most likely to fail.
In particular, the phenomenon of Barren Plateaus (BPs)~\cite{mcclean2018barren}, or exponentially vanishing gradients, even in its most benign form, greatly limits the extent to which quantum circuits can be trained at scale. BPs result from too expressive anstazes~\cite{mcclean2018barren,cerezo2021cost,holmes2021connecting,larocca2021diagnosing}, the usage of global cost functions~\cite{cerezo2021cost}, the generation of entanglement~\cite{sharma2020trainability,marrero2020entanglement}, or the presence of noise~\cite{wang2020noise,franca2020limitations}. 

Recall that BPs  are defined as an exponential decay of the magnitude of the circuit gradients with $n$. We denote by 
\begin{equation}\label{eqn_partial_i}
    \partial_i E \equiv \partial E(\thv) / \partial \thv_i
\end{equation}
the $i$-th element of the gradient vector of the cost $E(\thv)$ in \cref{eq:exp}, and by $\text{Var}( \partial_i E)$ the variance of such elements obtained over random circuit parameters. 
BPs occur when  $\text{Var}( \partial_i E)\in \OC(1/\rm{exp}(n))$, that is, when the partial derivatives concentrate exponentially in the system size. In practice, values of the cost and its gradients are estimated with an error $\sim 1/\sqrt{N}$ for a number of measurement repetitions $N$. 
Hence, in the case of BPs, $N$ would need to scale exponentially in order to distinguish true values of the gradients or cost function differences from statistical noise~\cite{arrasmith2020effect}. This would render optimization unfeasible beyond some critical system size. In the following we review how such BPs, particularly those caused by expressibility and those caused by noise, connect to the design of quantum circuits.

\subsubsection{Expressibility and the Dynamical Lie Algebra}\label{sec:expr}

Circuits acting on $n$ qubits realize unitaries belonging to $\mbb{U}(d)$, the group of unitaries of dimension $d$. For a given choice of circuit layout, it is of interest to understand what subset $U(\thv) \subseteq \mbb{U}(d)$ can be reached when varying the values of the parameters $\thv$, i.e., to understand how expressive a circuit is. To this intent, several metrics have been proposed~\cite{sukin2019expressibility,larocca2021diagnosing,larocca2021theory}. 
Here, we restrict our attention to the dimension of the circuit's \emph{Dynamical Lie Algebra} (DLA), a measure originating from the field of algebraic quantum optimal control~\cite{zeier2011symmetry} that was revisited in the context of VQAs~\cite{larocca2021diagnosing,larocca2021theory}.
We now briefly review such concept, and refer the reader to~\cite{larocca2021diagnosing} for a more detailed exposition. 

Starting from the generator set $\mathcal{G}$ of a given circuit, one can define its DLA (denoted $\mathfrak{g}$), as the vector space spanned by the individual generators and their repeated nested commutators:
\begin{equation}
 \mathfrak{g} = \text{Span} \langle i \mathcal{G} \rangle_{\rm{Lie}},
\end{equation}
where $\langle \cdot \rangle_{\rm{Lie}}$ denotes the Lie closure, the set generated by repeated nested commutators of the set. 
In turn, it can be shown~\cite{zeier2011symmetry} that unitaries realized by parametrized quantum circuits with a generator set $\mathcal{G}$, are of the form $U(\thv)=e^{g}$ for some $g \in \mathfrak{g}$. That is, the Lie algebra $\mathfrak{g}$ fully characterizes the (continuous Lie group of) unitaries that can be realized by the circuit.

Let us stress that the ability of $U(\thv)$ to realize exactly $U=e^{g}$, for any $g \in \mathfrak{g}$, assumes a sufficiently large number of layers.
Still, understanding circuits' expressibility through their DLA bears a particular appeal for our purposes.
First, great progress has been made in characterizing DLAs of some common circuit constructions~\cite{kazi2022landscape}. 
Second, it has recently been conjectured, and verified in several examples, that $\text{Var}( \partial_i E) \in \OC(1/ \rm{\poly}(\dimg))$~\cite{larocca2021diagnosing,zhang2021quantum}. 
In other words, BPs can arise as a consequence of a DLA that scales exponentially with $n$, as is the case of HEA circuits. 
Given such connection to circuit trainability, we will resort to DLA characterization when assessing expressibility of circuits later on.

\subsubsection{Noise}\label{sec:noise}

As a complementary aspect of trainability, noise has been recognized as a main obstacle when scaling quantum algorithms in the NISQ era~\cite{preskill2018quantum}. This was recently formalized in the context of VQAs~\cite{wang2020noise,franca2020limitations}.
Moreover, while error mitigation~\cite{temme2017error,kandala2018error,endo2021hybrid,czarnik2020error} can correct noisy observables, it does not necessarily solve noise-induced trainability issues since it also amplifies shot noise~\cite{wang2021can,takagi2021fundamental}.

This issue can already be appreciated for the most simple models of noise, such as a the global depolarization one defined as a channel $\Lambda(\rho)= p I/d + (1-p) \rho$ acting on (mixed) states $\rho\in \mathbb{C}^{d \times d}$,  with $d$ the dimension of the quantum state, $p\in[0,1]$ the depolarization probability, and $I$ the $d \times d$ identity.
Given a cost of the form $\Tr[\ham\rho]$ (generalizing~\cref{eq:exp} to mixed states), and assuming that the same channel acts at each of the $L$ layers of a circuit, one can see that the noisy cost function decays as $(1-p)^{L} \Tr[\ham \rho]$ (assuming a traceless $\ham$), i.e., exponentially with the number of layers. 
Hence, akin to the BPs previously discussed, one would need to commit a number of measurements exponentially growing with (this time) the depth of the circuits employed, thus rendering the optimization of deep circuits impractical. 
While discussed here for simplistic noise, more involved and realistic noise models, including local Pauli noise channels, have been proven to yield similar effects~\cite{wang2020noise}.

\medskip
Given such limitations, it is highly desirable to design circuits that could approximate the target solution in with a  depth that is a shallow as possible.
In this respect, HEA circuits seem naturally suited, as they provide freedom in the choices of parameter values and enhanced expressibility for a given number of layers, but unfortunately they have expressibility-induced BPs when enough layers are used~\cite{mcclean2018barren,cerezo2020cost}. 
On the other hand, HVA--like ansatzes  have been shown to reduce BPs problems~\cite{wiersema2020exploring}, but due to their rigid parameter structure seem to require unnecessarily deep circuits.   
In the next section, we discuss how symmetry principles provide a way forward in the design of quantum circuits that generalize HVA--like ansatz constructions, so as to reconcile, to some extent, these two conflicting aspects.

\section{Orb circuits}\label{sec:orb}

To motivate the construction of \orbitcircuit s, we start by recalling relevant aspects of symmetries in quantum systems (\cref{sec:sym}). This understanding can be used to reduce the dimensionality of the space of states that need to be explored, or equivalentely to reduce the expressibility of the circuits that need to be used, and thus to ease the present tasks of ground state preparations. Then, we recast such restricted exploration in terms of requirement of circuit properties, namely \emph{equivariance} under the relevant group of symmetries (\cref{sec:autinv}). As we will see, such requirements are fulfilled with HVA-like circuits, but, at the cost of often too much restrictions. In particular, HVA-like circuits greatly limit the number of independent parameters per circuit layer. This hints at the fact that they would require deeper circuits than is necessary. 
This will lead us to the development of \orbitcircuit s (\cref{sec:orb}) generalizing and improving on HVA-like circuits, in the sense that they also respect the relevant symmetries of the Hamiltonian, but can allow for much more independent parameters.

\subsection{Symmetries}\label{sec:sym}
A symmetry $S$ of a Hamiltonian $\ham$ is a unitary operator ($SS^{\dag}=S^{\dag}S=I$) leaving $H$ invariant, such that $S \ham S^{-1}=\ham$ (or equivalently, such that $[S, \ham]=0$). 
In cases where $\ham$ admits more than one symmetry (the identity operation being a trivial one), these organize as groups. In particular, given two symmetries $S_1$ and $S_2$ their compositions ($S_1S_2$ or $S_2S_1$) and their inverses ($S_1^{-1}$ and $S_2^{-1}$) are also symmetries.

Symmetries of quantum systems can be broadly classified in two classes. On one hand, spatial symmetries (or outer symmetries~\cite{zeier2011symmetry}) are those symmetries which are permutations of sites of the system.
These form a discrete group of symmetries, with order as large as $n!$ for the case of the \emph{Symmetric group} $S_n$, and will be the symmetries of particular interest for this work.
Not all symmetries fall in this first category and the remaining ones are termed internal symmetries (or inner ones).
These can either be discrete groups, as is the case of $\mathbb{Z}_2=\set{I, X^{\otimes n}}$, or continuous ones, as is the case of $\mathbb{SU}_2=\set{U^{\otimes n}|U\in \text{SU(2)}}$. Designing quantum circuits that respect such inner symmetries is usually achieved by a choice of gate generators commuting with them, e.g., $\mathcal{G}_1=\set{\swap_{i,j}}$ with $\swap_{i,j}$ the swap operator between qubit $i$ and $j$ for the case of $\mathbb{SU}_2$.

Turning back our focus to spatial symmetries, and considering a $n$--qubit system, we denote by $\pi \in S_n$ the permutations over the set of $n$ indices that label the qubits. The action of $\pi$ on a n--qubit state $\ket{\psi} \rightarrow \permutestate{\psi}{\pi}$ is understood as the corresponding permutation of its qubit indices (note the overloading of the symbol $\pi$ to a unitary operator when acting on states). For instance, for a state $\ket{\psi} = \ket{s_1}\ket{s_2}\ket{s_3}$ and $\pi(1)=2$, $\pi(2)=1$ and $\pi(3)=3$ (i.e., a permutation of the indices $1$ and $2$), then 
$\pi \ket{\psi} = \ket{s_2}\ket{s_1}\ket{s_3}$.  
Accordingly, $\pi$ acts on operators as $O \rightarrow \permuteop{O}{\pi}$.

We call the spatial symmetries of a Hamiltonian its automorphism group, that is defined as follows.  
\begin{definition}\label{def:auto}
Given a Hamiltonian $H$, we define its automorphism group $\aut \subset S_n$ as the subgroup of permutations $\pi_a$ that leave $\ham$ unchanged. 
\begin{equation}\label{eq:auto}
\aut = \set{\pi_a \in S_n | \permuteop{\ham}{\pi_a}=\ham}.
\end{equation}
\end{definition}

Given commutation of $\ham$ with an element $\pi_a \in \aut$, it is known that $\ham$ can be simultaneously diagonalized with $\pi_a$, that is, $\ham$ is block diagonal in the eigenbasis of $\pi_a$. 
More generally, operators $\ham$ commuting with groups of unitaries, as in \cref{eq:auto}, will also adopt a block decomposition inherited from the structure of the group.
For instance, in the case of translational invariance of a system, each of these blocks corresponds to a subspace of states with given momentum.
Crucially, the search for a ground state can be limited to a search within each (or a subset) of the individual symmetry blocks, thus reducing the Hilbert space that has to be considered and the expressibility of the circuits employed.
Given that circuits with the same symmetries also share the same block structure, when applied to an initial state with support in a single block, one can ensure that these will prepare states remaining in  the same block, thus allowing exploration of a smaller subspace of the Hilbert space.

In the following, we focus our attention on ground-state preparations in the \emph{invariant subspace} of $\aut$ which contains all the states $\ket{\psi(\thv)}$ such that for all $\pi_a  \in \aut$, $\ket{\psi(\thv)}=\pi_a \ket{\psi(\thv)}$. 
As we now discuss, search in this subspace can be efficiently parameterized by equivariant circuits applied to easy-to-prepare initial states such as $\ket{0}^{\otimes n}$ or $\ket{+}^{\otimes n}$. Note however that the same techniques are directly applicable to any other block defined by $\aut$, as long as the corresponding initial states can be efficiently prepared.

\subsection{Automorphism-Invariant state preparation}\label{sec:autinv}

Here, we show that given  (i) an initial state invariant under permutations, and (ii) a circuit equivariant with respect to $\aut$ one can ensure that the states prepared by the quantum circuit are invariant under $\aut$.  

\begin{definition}\label{def:autoinv}
A circuit $U(\thv)$ is said to be equivariant with respect to a symmetry group $\mathbb{G} \subset S_n$ whenever 
\begin{equation}\label{eq:equiv_circuit}
    \forall \pi \in \mathbb{G},\;  U(\thv) = \permuteop{U(\thv)}{\pi}\ .
\end{equation}
We denote the space of such circuits as $\mathcal{E}(\mathbb{G})$. 
\end{definition}
Given $U(\thv) \in \autinv$ and a symmetric input state $\ket{\psi_0}$ (e.g., $\ket{0}^{\otimes n}$ or $\ket{+}^{\otimes n}$), it can be verified that, as desired, the realized state is invariant under any $\pi_a \in \aut$:
\begin{align}
\begin{split}
     \ket{\psi(\thv)} &= U(\thv) \ket{\psi_0} = \permuteop{U(\thv)}{\pi_a}\ket{\psi_0} \\
     & \pi_a U(\thv)\ket{\psi_0} = \permutestate{\psi(\thv)}{\pi_a}.
\end{split}
\end{align}
Note that the requirement of a symmetric initial state could be relaxed to a requirement of an initial state invariant under the automorphism group only.
However, this may result in input states that are hard to prepare in practice.

The equivariance property of unitaries as defined in \cref{def:autoinv} is associative by composition. That is, given $U_1$ and $U_2 \in \autinv$, their product, $U_1U_2$ or $U_2U_1$, also belongs to $\autinv$). 
As such, imposing equivariance of a circuit can be achieved in a constructive manner by imposing equivariance at its constituent level, i.e., at the level of its layers.

As we show in the next section, HVA-like circuits naturally enforce such equivariance properties, thus effectively limiting  their expressibility and improving their trainability. 
This, however, is achieved at the cost of reducing greatly the number of independent parameters per layer. 
Still, they can be readily generalized to \orbitcircuit s that preserve equivariance, but also, that maximize the number of free parameters per layer.

\subsection{From HVA to ORB circuits}\label{sec:orblc}

Let us begin by discussing HVA-like circuits. As mentioned earlier, and exemplified in~\cref{eq:alt}, HVA-like circuits are most often composed of $1$- and $2$-qubit gates with shared parameters.

\begin{figure}
	\includegraphics[width=0.46\textwidth]{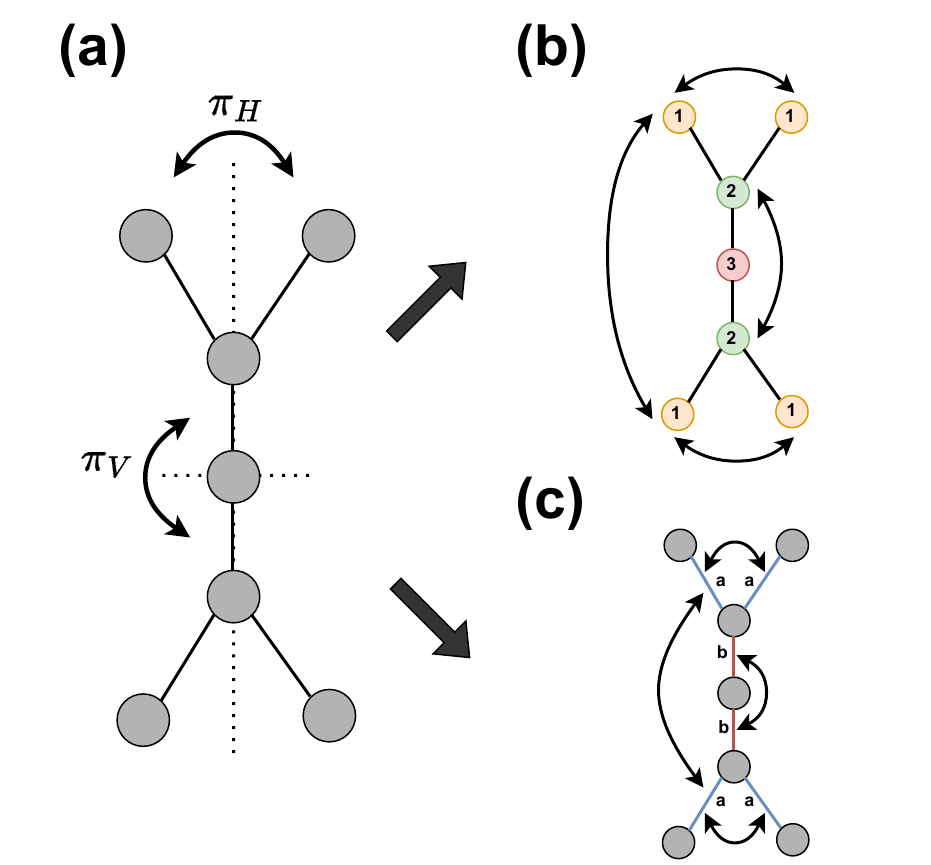}
	\caption{\textbf{Constructions of orbits and edge-orbits.}
    (a) Consider a system with automorphism group $\aut=\set{1, \pi_H, \pi_V, \pi_H \pi_V}$ generated by the horizontal and vertical reflections, $\pi_H$ and $\pi_V$ respectively.
    (b) According to these symmetries, the different parts of the systems group as $3$ distinct orbits (numbered from $1$ to $3$) as per \cref{def:orbit}. Any two parts inter-orbit can be related by an element of $\aut$, while any two parts intra-orbit cannot.   
    (c) Similarly, the edges group as $2$ edge-orbits (labeled a and b) as per \cref{def:eorbit}.
	}
	\label{fig:orbits}
\end{figure}

\subsubsection{Layers of $1$-qubit gates}

Focusing first on the $1$--qubit gates, the layer in~\cref{eq:alt} is composed of a sublayer, 
\begin{equation}\label{eq:Vx}
    V_X(\theta)= \prod_{i=1}^n R_{X_i}(\theta),
\end{equation}
of X rotations with the same angle $\theta$. Since $\pi V_X(\theta) \pi^{-1} = V_X(\theta)$ for \emph{any} permutation $\pi \in S_n$, we have  $V_X(\theta) \in \autinv$ for \emph{any} $\ham$. However, this comes at the expense of imposing the same rotation to be applied to each of the qubits, and seems poorly tailored to the exact details of $\ham$.

A key insight is to realize that while it is reasonable to ask that \emph{equivalent} qubits, with respect to $\ham$, should experience similar transformations, this does not have to extend to all qubits. To formalize further this notion of equivalence we introduce the concepts of orbits.

\begin{definition}[Orbit of \ham]\label{def:orbit}
An orbit of a Hamiltonian $\ham$ acting on a system of $n$ qubits is defined as a set of qubit indices closed under the action of \aut. 
The orbit of an arbitrary qubit--index $i$ is given by 
\begin{equation}\label{eq:vertex_orb}
    \orbit_{i} = \set{\pi_a(i)}_{\pi_a \in \aut}\,.   
\end{equation}
\end{definition}
When $j\in \orbit_{i}$, both the orbits $\orbit_i$ and $\orbit_j$ contain the exact same indices, and are thus deemed equivalent. 
We label non-equivalent orbits as $\orbit_o$ with a separate index $o$ taking values in a subset of $\set{1,\hdots, n}$. Also, we denote the set of all distinct orbits $\orbit = \set{\orbit_o}$, which has cardinatlity $|\orbit|$ and forms a partition of the set of all qubit indices. 
Construction of orbits are illustrated in~\cref{fig:orbits} for a system admitting $2$ reflection symmetries, resulting in $|\mathcal{O}|=3$ orbits numbered as $1$, $2$ and $3$ in~\cref{fig:orbits}(b).

The ORB generalization of $V_X(\theta)$ in~\cref{eq:Vx} is defined as:
\begin{equation}\label{eq:xorb}
    V_X^{ORB} (\thv) = \prod^{|\mathcal{O}|}_{o = 1} \prod_{i \in \mathcal{O}_o} R_{X_i} (\thv_o),
\end{equation}
which applies the \emph{same} rotation to any qubit within the \emph{same} orbit, but allows for \emph{different} rotations in between \emph{distinct} orbits.
One can readily verify that any further division of this parameter grouping  (i.e., the addition of any extra free parameter) would break the equivariance of the layer. That is, the parameter structure in~\cref{eq:xorb} maximizes the number of independent parameters while still maintaining equivariance of the layer.

In the case when $\aut=\{I\}$, each qubit belongs to its own orbit so that each rotation is assigned a distinct angle, yielding $|\orbit|=n$ parameters per layer. When $\aut=S_n$, each qubit belongs to the same orbit and $|\orbit|=1$.  In general, $V^{ORB}_X$ packs a number of parameters $|\mathcal{O}|$ in between $1$ and $n$, that is the maximum possible (for a given $\ham$) while still preserving equivariance.
While exemplified here for $X$-rotations, this construction equally applies to any layer of $1$-qubit rotations. 

\subsubsection{Layers of $2$-qubit gates}
The concept of orbits can be extended to orbits over pairs of qubits (called edges).
\begin{definition}[Edge-orbit of \ham]\label{def:eorbit}
An edge-orbit of a Hamiltonian $\ham$ acting on a system of $n$ qubits is defined as a set of edges that remain closed under the action of \aut.  The orbit generated by an arbitrary edge $(i,j)$ is given by
\begin{equation}\label{eq:orbit}
    \eorbit_{(i,j)} = \set{\big(\pi_a(i),\pi_a(j)\big)}_{\pi_a \in \aut}\,.   
\end{equation}
\end{definition}
As for orbits, any two edge-orbits are either exactly the same or non-intersecting, and the set of distinct edge-orbits $\eorbit=\set{\eorbit_{m}}$ partitions the full set of edges. 
We can further specialize this definition to edges in a subset $\mathcal{I}$ of all possible edges corresponding to the physical interactions of the system, that is, which appear as two--body terms in $\ham$.
It can be verified that such restriction is consistent, i.e., given an edge $(i,j)$ appearing in $\ham$ any of its transform $(\pi_a(i), \pi_a(j))$ for $\pi_a \in \aut$ will also appear in $\ham$ (by definition of $\aut$). Construction of edge-orbits are illustrated in~\cref{fig:orbits}(c) for the example mentioned earlier.   

As before, we can generalize layers of $2$--qubit gates appearing in HVA-like circuits by grouping parameters per edge-orbits. 
For instance, the layer of $2$-qubit rotations $V_{ZZ}(\theta)= \prod_{(i,j) \in \mathcal{I}} R_{Z_i Z_j}(\theta)$ appearing in~\cref{eq:alt} becomes:
\begin{equation}\label{eq:zzorb}
    V_{ZZ}^{ORB} (\thv) = \prod^{|\mathcal{O}^e|}_{m = 1} \prod_{(i,j) \in \mathcal{O}^e_m} R_{Z_iZ_j} (\thv_m),
\end{equation}
that could pack up to $n(n-1)/2$ distinct parameters, as opposed to a single parameter for HVA-like circuits, while still retaining equivariance under $\aut$. As was the case for \cref{eq:xorb}, one can verify that any additional independent parameter would break equivariance of \cref{eq:zzorb}.

This equivariant construction can be extended to any set of $2$--qubit (and in principle to higher order) interactions as 
\begin{equation}\label{eq:genorb}
    V_{P}^{ORB} (\thv) = \prod^{|\mathcal{O}^e|}_{m = 1} \text{exp}\Big[- i \thv_m \sum_{(i,j) \in \mathcal{O}^e_m} P_{i,j}\Big],
\end{equation}
where $P$ denote the type of interactions involved in between pairs of qubits. 

\begin{figure*}
	\includegraphics[width=0.90\textwidth]{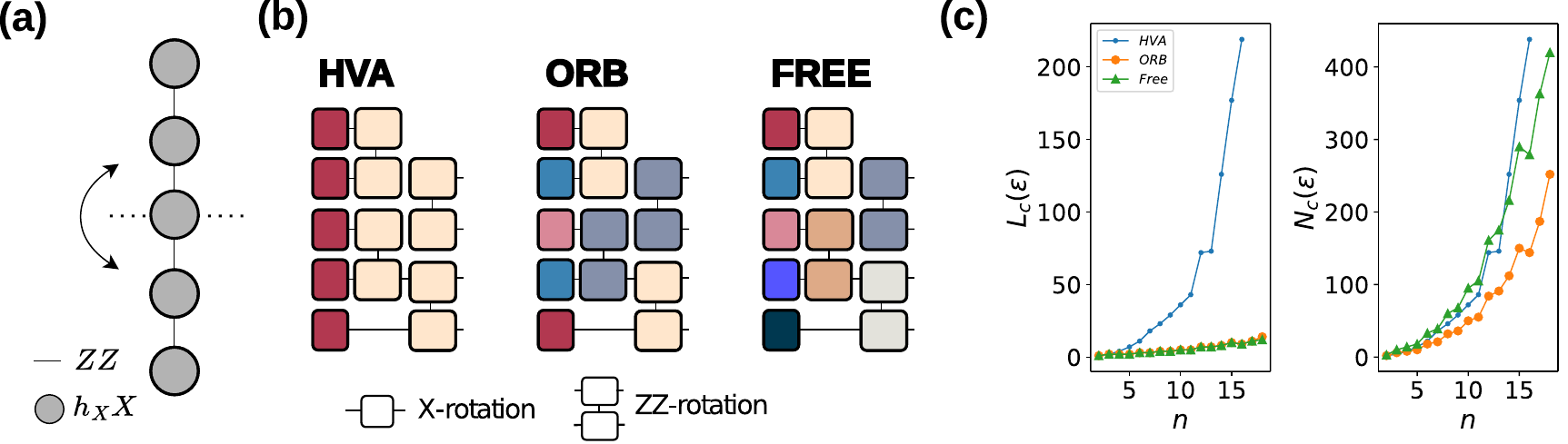}
	\caption{
    \textbf{Transverse--field Ising model (TFIM) in one-dimension with open boundary conditions.} 
    (a) The Hamiltonian defined in \cref{eq:Htfim} is graphically depicted here for $n=5$ spins. 
    Each qubit interacts with its nearest-neighbors (ZZ interactions) and subjected to a transverse X field of global strength $h_X$. 
    The reflection w.r.t.\ the middle of chain is the only spatial symmetry of the Hamiltonian.     (b) Three different circuit constructions are studied, all involving parameterized $1$-qubit X rotations and $2$-qubit ZZ rotations but with different number $n_l$ of \emph{independent} parameters (corresponding to \emph{distinct} colors) per layer, increasing from left to right.  
    For the HVA the parameters are tied by gate type with $n_l=2$  parameters per layer.
    In the case of \orbitcircuit s parameters are tied both by gate-type and also orbits (\cref{def:orbit,def:eorbit}) yielding $n_l=5$ parameters. 
    Lastly, each gate in the Free layer is assigned an independent parameter with $n_L=9$ parameters. 
    (c) Performance of the circuits for system sizes $n \in [2,18]$. 
    For each circuit (colors in legend), results are reported in terms of the critical number of layers ($L_c(\varepsilon)$ on the left) and independent parameters ($N_c(\varepsilon)$ on the right) necessitated to achieve a relative error $r$ smaller than  $\varepsilon =10^{-5}$.  
	}
	\label{fig:tfim1d}
\end{figure*}

When any of the $P_{i,j}$ interactions intra-orbit commute, \cref{eq:genorb} is readily implemented as a composition of $2$-qubit gates $R_{P_{i,j}}(\thv_m)= \text{exp}[-i \thv_m P_{i,j}]$.
This was the case for $P=ZZ$, resulting in~\cref{eq:zzorb}.
However, cases where interaction terms do not commute within orbits require more care, and will be detailed later in~\cref{sec:J1J2}. 
A simple example of application of \orbitcircuit s with layers of the form~\cref{eq:xorb,eq:zzorb} is illustrated in~\cref{fig:tfim1d} and discussed in the next section.

\section{Numerical Results}\label{sec:res}
Having described the general constructions of \orbitcircuit s, we proceed by studying their performance on several benchmark problems. 
These benchmarks are chosen to span a variety of physical models (with local and non-local interactions) and underlying spatial topologies ($1$D and $2$D lattices, and also random connectivity). 
\Cref{sec:meth} reviews the methodology adopted for these studies.
Then, \orbitcircuit\ is compared to other circuit constructions in $4$ different examples: the transverse-field Ising model (TFIM) in $1$D (\cref{sec:tfim1d}) and $2$D (\cref{sec:tfim2d}), problems of Max-Cut on random graphs (\cref{sec:mc}) and the $J_1$--$J_2$ Heisenberg model in $2$D (\cref{sec:J1J2}).

\subsection{Methodology}\label{sec:meth}
In the following, we aim at studying the performances of \orbitcircuit s -- especially their scaling with respect to increased problem sizes -- and at comparing them to other related ansatzes. We now review how these different aspects are probed.

The performance of a given circuit, at fixed number of layers $L$, can be assessed in terms of the energy $E^{opt}(L)$ achieved after parameter optimization, relative to the true ground-state energy $E_{GS}$ of the Hamiltonian under study.
This is quantified in terms of the relative error
\begin{equation}\label{eq:ratio}
    r(L) = \frac{E^{opt}(L)-E_{GS}}{|E_{GS}|}.
\end{equation}

To gauge the performance of a circuit construction across different system sizes, we will report the critical number $L_c(\varepsilon)$ of layers required to achieve a sufficiently small error $\varepsilon$.
For each family of circuits and each system size $n$ considered, this number is obtained by repeating optimizations of circuits with increasing number of layers $L$ until $r(L) \leq \varepsilon$.
Additionally, we will report the total number $N_c(\varepsilon) = n_l \times L_c(\varepsilon)$ of independent parameters corresponding to $L_c(\varepsilon)$ layers and where $n_l$ denotes the number of independent parameters per layer (that varies depending on the family of circuits considered).

All the circuits are numerically simulated using Tensorflow Quantum~\cite{broughton2020tensorflow} and their optimizations are performed using the L-BFGS-B routine~\cite{zhu1997algorithm} with initial random circuit parameters and a number of iterations systematically capped to $500$ iterations per individual optimization. Given such stochastic initialization, any optimization is repeated $N_{rep}=25$ times and results are reported in terms of the statistics (median) evaluated over these repetitions.

These optimization metrics already capture some main aspects of trainability. Namely, non-vanishing relative errors arise from limited expressibility of the circuits, and also from features of the optimization landscape such as the presence of low-quality local minima. Still, given that these figures are obtained on simulations, they do not reflect problems of vanishing gradients. 
Hence, we will also provide an assessment of the scaling of the variance of the parameter gradients. Variances $\text{Var} (\partial_i E)$ of the individual partial derivatives (see Eq.~\eqref{eqn_partial_i}) 
are evaluated over repeated random circuit initializations, and depend both on a given parameter index $i$ and on the depth $L$ of the circuit considered.
To provide synthetic but meaningful information, we will report the median \begin{equation}\label{eq:grads}
    \text{Var}_{c}(\varepsilon)  \equiv  \langle \text{Var}(\partial_i E)\rangle_i \text{, for } L = L_c(\varepsilon)
\end{equation}
evaluated at a critical number of layers, that is, in a regime where high-fidelity ground state preparation is in principle possible. 

\subsection{One-dimensional transverse--field Ising model}\label{sec:tfim1d}
As a starting example, we consider the Transverse-Field Ising Model (TFIM) with Hamiltonian
\begin{equation}\label{eq:Htfim}
    \ham_{\text{TFIM}} = - \sum_{\langle i, j\rangle} Z_{i}Z_{j} - h_x \sum^n_{i=1} X_i
\end{equation} 
consisting of $Z_{i} Z_{j}$ interactions between neighboring pairs of qubits $\langle i,j \rangle$, and a global transverse field $\sum_i X_i$ affecting equally each of the $n$ qubits with strength $h_x>0$. 

In $1$D, qubits are arranged over a line such that a qubit $i \in [2, n-1]$ has neighbors $i\pm1$. 
Additionally, for periodic boundary condition (i.e., a ring) qubits $1$ and $n$ are taken to be neighbors, while for open-boundary conditions (i.e., a path) both qubits $1$ and $n$ have a single neighbor. This $1$D model is known to exhibit a transition from a ferromagnetic phase ($h_x<1$) to a disordered paramagnetic one ($h_x>1$), and can be exactly solved by means of the Jordan--Wigner transformation. Such minimal model of a quantum phase transition has been extensively used as a prototype of quantum magnetism and as a testbed for the assessment of VQA performances; it will serve us as a first benchmark. 

In Refs.~\cite{ho2019efficient,wiersema2020exploring}, high--fidelity preparation of the ground state of the TFIM on a ring is consistently achieved using HVA circuits with critical depth and number of parameters growing linearly with the system size. The circuits employed have layers of the form~\cref{eq:alt} 
applied to an initial state $\ket{\psi_0}=\ket{+}^{\otimes n}$.  
In this case, the Hamiltonian remains invariant under any translation $\mathcal{T}_k(i)=i+k\, (\text{mod } n)$ of the qubits by a number of sites $k$. Given the $1$D nature of the system, any qubit $j$ can be mapped to any other qubit $j'$ with $\mathcal{T}_{j' - j}$. 
In other words, each qubit (and edge) is equivalent to another, thus yielding a single orbit (and a single edge--orbit) as per \cref{def:orbit,def:eorbit}; in this case \orbitcircuit\ effectively recovers the HVA ansatz.

The situation differs when studying the same system with open boundary conditions, which is depicted in~\cref{fig:tfim1d}(a). Now, the Hamiltonian only admits a single non-trivial automorphism $\mathcal{R}(i)=n+1-i$, that permutes qubits from each side of the chain. 
Accordingly, qubits group as $|\mathcal{O}|=n/2$ ($|\mathcal{O}|=(n+1)/2$) orbits, for even (odd) values of $n$. 
Each orbit consists in either a pair of qubit indices $(i,n+1-i)$ or a single one when $i=(n+1)/2$.
Similarly, edges groups as $|\mathcal{O}^e|=n/2$ ($|\mathcal{O}^e|=(n-1)/2$) edge-orbits for even (odd) $n$. 
Consequently, an \orbitcircuit\ allows for $n_l=n$ free parameters per layer, a figure to be compared to the $n_l=2$ parameters for the HVA.

Both a layer of HVA and an \orbitcircuit\ are depicted in \Cref{fig:tfim1d}~(b) for a system size of $n=5$ sites.
For each ansatz, independent parameters are displayed as distinct colors. For benchmarking purposes, we also include a ``Free'' circuit, that consists of the same set of gates as the HVA and \orbitcircuit s, but with each gate being assigned an independent parameter yielding a total of $n_l=2n-1$ parameters per layer. 

Naively, one would have expected the expressibility of the circuits to increase monotonically with the number $n_l$ of free parameters per layer. 
However, it can be shown~\cite{kazi2022landscape} that both \orbitcircuit\ and HVA have the same expressibility, growing quadratically with the system-size. 
In other words, despite the possibility of packing $n/2$ times more parameters per layer, \orbitcircuit\ spans (in the limit of large $n$) exactly the same group of unitaries as its HVA alter-ego.
In contrast, the ``Free'' ansatz has increased expressibility~\cite{kazi2022landscape}, and overall we have $\text{dim}(\mathfrak{g}_{Free}) > \text{dim}(\mathfrak{g}_{ORB}) = \text{dim}(\mathfrak{g}_{HVA})=n^2$. 
We now probe further how such appealing feature translates into improved performances.

\Cref{fig:tfim1d}~(c) reports the critical number $L_c(\varepsilon)$ of layers (left panel), and critical number  $N_c(\varepsilon)$ of parameters (right panel), required to  achieve errors as low as $\varepsilon=10^{-5}$.
As can be seen, both \orbitcircuit\ and the ``Free'' construction requires a number of layers substantially smaller than HVA. Already for system sizes of $n = 15$ qubits this corresponds to more than an order of magnitude reduction in the number of layers required, a significant edge in light of current noise levels of quantum hardware (e.g., see Ref.~\cite{wang2020noise}). 

Furthermore, \orbitcircuit\ is found to necessitate the least number of parameters to achieve a similar performance when compared to the other circuits studied. 
Note that the observed quadratic scaling in the number of parameters matches closely the dimension of the DLAs consistently with Ref.~\cite{larocca2021theory}. 
Overall, in this example, \orbitcircuit\ emerges as an optimal construction, in terms of number layers and parameters required, allowing us to pack as many parameters per layer as is possible without increasing the expressibility of the ansatz.

\subsection{Two-dimensional transverse-field Ising model}\label{sec:tfim2d}
\begin{figure}[t!]
	\includegraphics[width=0.48\textwidth]{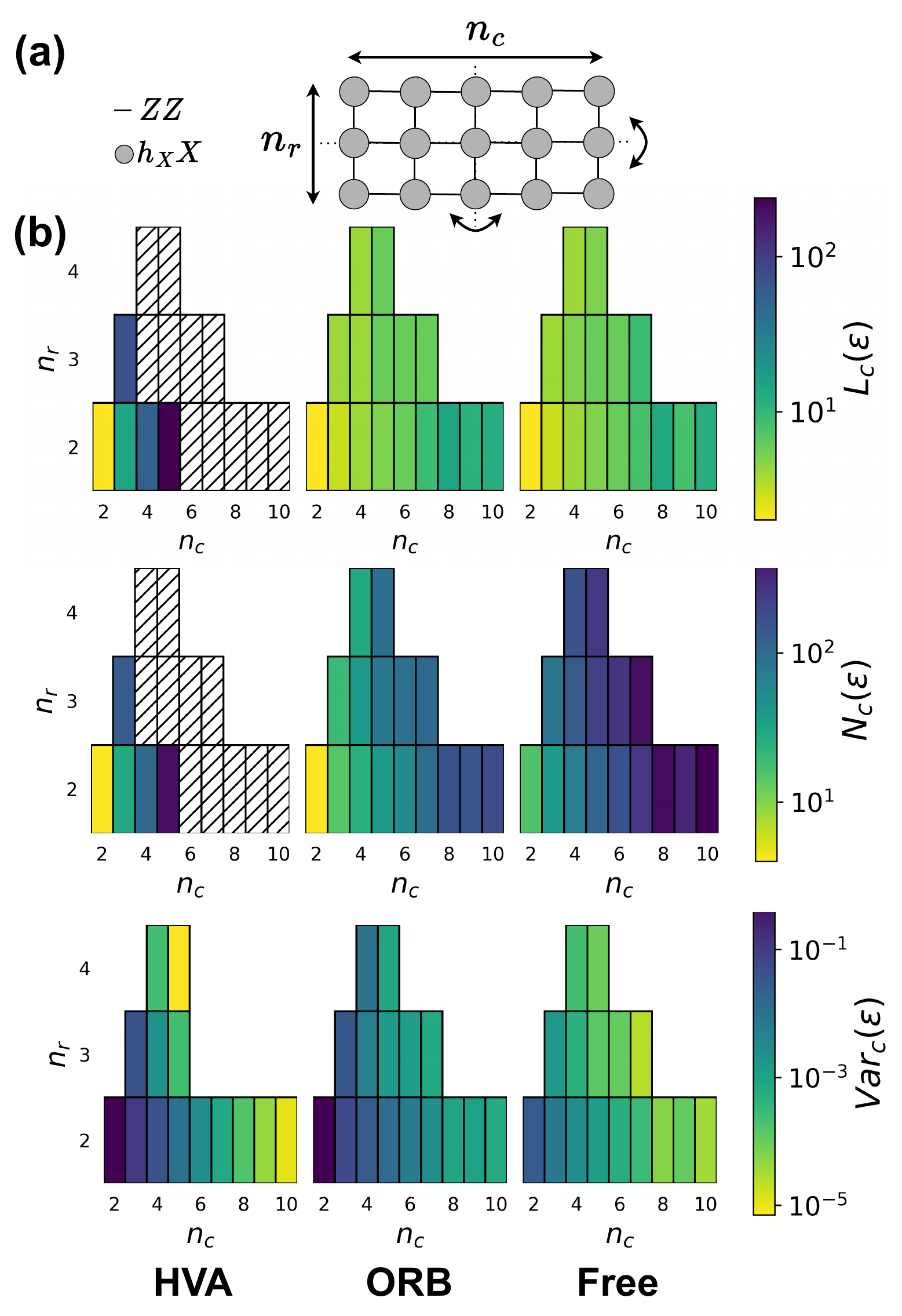}
	\caption{
    \textbf{TFIM in $2$D.} 
    (a) The TFIM model in~\cref{eq:Htfim} is studied on a $n_r \times n_c$ grid. 
    (b) Performances of the circuits (HVA, ORB, and Free) are shown for $n_c\in[2,7]$ and $n_r \in [2,4]$. These are reported in terms of the critical number of layers (first row), parameters (mid row) and variance of the gradients (defined in~\cref{eq:grads}, last row) for a relative error $\varepsilon=10^{-3}$. 
    Cases where an anstaz did not achieve such error, within $L \leq 500$ layers, are reported as hachure.
	}
	\label{fig:tfim2d}
\end{figure}

While $1$D problems already offer invaluable insights, the most compelling, and challenging, physical models often arise in higher dimensions.
We now study further applications of \orbitcircuit\ to the $2$D TFIM laid over regular grids composed of $n_r \in [2,4]$ rows and $n_c \in [2, 7]$ columns involving up to $n=21$ qubits (depicted in~\cref{fig:tfim1d}(a)). 

The ansatzes studied earlier directly extend to this $2$D scenario. The sublayer of interactions $V_{ZZ}(\theta)=\sum_{\langle i,j\rangle} R_{Z_iZ_j}(\theta)$, appearing in an HVA ansatz, now runs over all the pairs of adjacent grid qubits $\langle i,j \rangle$, but is still parameterized by a single parameter. 
Additionally, $X$ rotations are applied to each of the $n$ qubits with the same angle per layer.
In the case of \orbitcircuit s these $X$ ($ZZ$) rotations are grouped by orbits (edge-orbits) resulting from the two reflections (one horizontal and one vertical) and also, in the case of $n_r=n_c$, from two additional ones (w.r.t. the diagonals). 
In the case of the Free ansatz, each gate corresponds to a distinct parameter.

Performances of the circuits for all the system sizes considered are depicted in \cref{fig:tfim2d}(b) including the critical number of layers (top row), number of parameters (middle row) and gradients variances (bottom row) for a relative ground-state error of $\varepsilon=10^{-3}$.
In line with previous results, we find \orbitcircuit\ to exhibit enhanced performances.
For the range of layers considered ($L\leq 500$), HVA circuits fail in converging for problems involving more than $10$ qubits. 
In contrast both the Free and \orbitcircuit\ achieve the desired accuracy in $L_c(\varepsilon)\leq 12$ layers for all the system sizes probed.
Furthermore, \orbitcircuit s necessitate a number of parameters significantly smaller than the Free ansatzes, with $4$ to $5$ times less parameters for the largest system studied.

We also compare the variance of the circuit gradients, defined in~\cref{eq:grads}, for the three ansatzes. 
Except for the smallest sizes studied, the gradients of \orbitcircuit s are found to be systematically larger than others. 
On one hand, HVA and \orbitcircuit s  share the same expressibility~\cite{kazi2022landscape}, such that at similar circuit depth one would expect similar gradient magnitudes. 
However, for a given error $\varepsilon$, \orbitcircuit s require significantly less layers than HVA such that its gradients remain larger. 
On the other hand, both ORB and Free circuits converge in a similar number of layers, but the expressibility of Free circuits~\cite{kazi2022landscape} is the largest, hence yield smaller gradients. 
Overall, these combined effects result in gradients of \orbitcircuit s that are one or two orders of magnitude larger than others. (This discussion does not account for noise-induced gradient suppression~\cite{wang2020noise}. Accounting for noise would lead to an even more dramatic separation between ORB and HVA, since HVA requires much deeper circuits than ORB.)

\begin{figure}
	\includegraphics[width=0.48\textwidth]{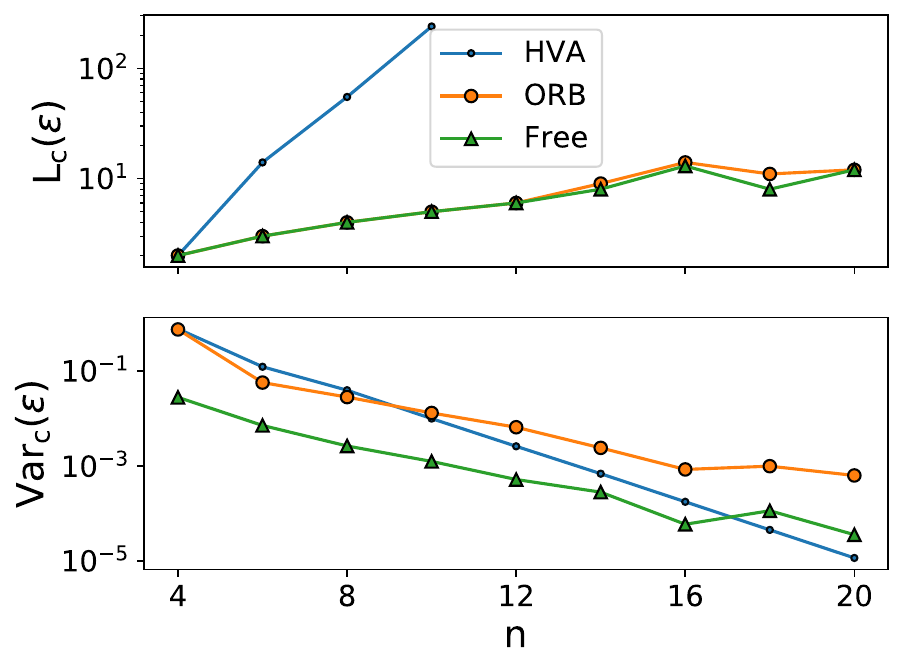}
	\caption{\textbf{2D TFIM scaling.} Results are shown for the TFIM on a $2\times n_c$ grid. We plot the critical number of layers (top row) and variance of the gradients (bottom row) as a function of the system size $n=2 \times n_c$.
	}
	\label{fig:tfim2d_grads}
\end{figure}

Such substantial improvement offered by \orbitcircuit s should not obscure the fact that the dimension of the DLAs of any of the circuits grows exponentially as a function of $n$~\cite{kazi2022landscape}, hinting to issues  of trainablity for too large system sizes~\cite{larocca2021diagnosing}. 
To probe further the scaling of the circuits properties, we focus on the case of the $2\times n_c$ grids, and report in Fig.~\ref{fig:tfim2d_grads} the critical number of layers (top row) and gradients variances (bottom row) as a function of the system size $n=2\times n_c$. 
As can be seen, the number of layers required to achieve precise ground-state preparation displays an exponential behavior over most of the system sizes studied, albeit at a much slower rate for ORB and Free circuits.
Concurrently, magnitudes of the gradients decay exponentially fast over these system sizes.
Still, we note a flattening of these exponential increases (in the number of layers) and decays (in the gradients magnitudes) for $n\geq16$ especially for ORB circuits. Notably, for $n>10$ qubits, the gradients of \orbitcircuit s are found to be significantly larger than for the HVA and Free circuits. That is, while not generically solving problems of BPs at large $n$, \orbitcircuit s are found to mitigate them.

\subsection{Max-Cut}\label{sec:mc}
\begin{figure}[t!]
	\includegraphics[width=0.48\textwidth]{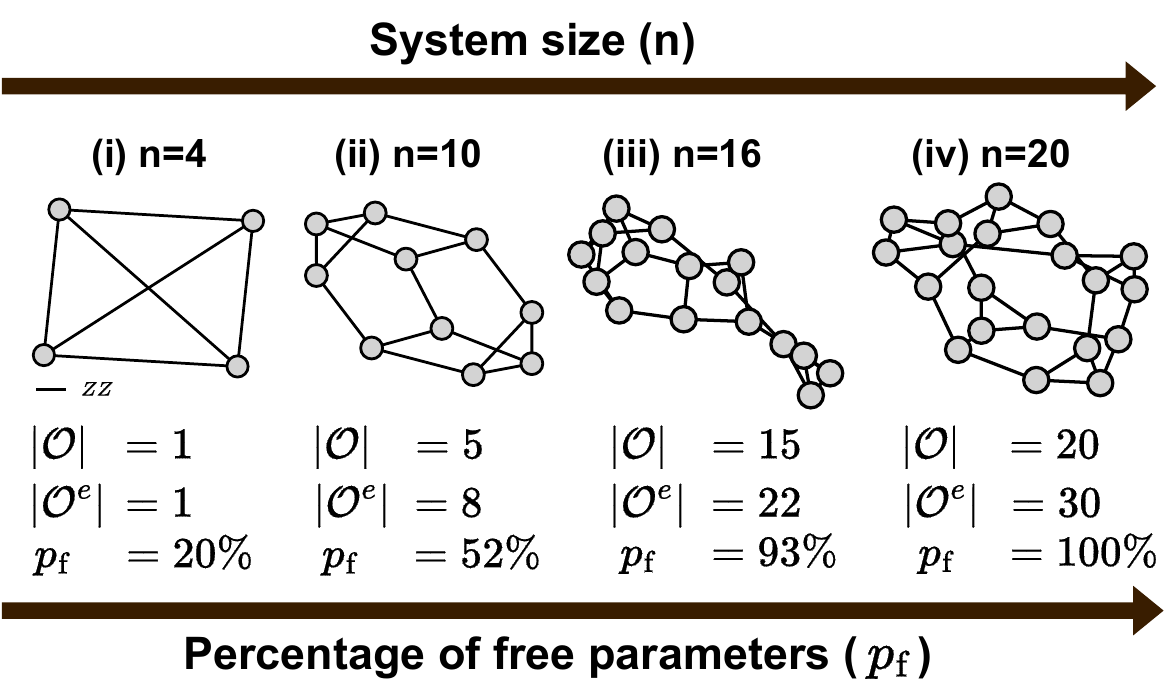}
	\caption{
    \textbf{Random $3$--regular graphs.} 
	Examples of random $3$--regular graphs for $n=4$ to $20$ nodes. 
	For each graph, we display the number of orbits and edge-orbits, denoted $|\mathcal{O}|$ and $|\mathcal{O}^e|$ respectively.
	Taken together, these give the number of parameters per layer of \orbitcircuit , $n_l=|\mathcal{O}|+|\mathcal{O}^e|$, and we report the percentage $p_\text{f}$ of such free parameters compared to the total number of gates.
	As the system size grows, this percentage quickly saturates to its maximal value of $100\%$ (i.e., each gate is independently parameterized), in stark contrast with QAOA circuit constructions. }
	\label{fig:mc_graphs}
\end{figure}

Having demonstrated the benefits of \orbitcircuit\ for physical models defined on regular lattices, we now aim at understanding its appeal when dealing with less-to-unstructured Hamiltonian topologies.
This situation is best exemplified for problems of Maximum-Cut (Max-Cut) addressed by means of the QAOA~\cite{farhi2014quantum}. 

\subsubsection{QAOA for Max-Cut}

Recall that for a given graph $\graph$, with set of n vertices $\mathcal{V}(\graph)$ and edges $\mathcal{E}(\graph)$, a Max-Cut Hamiltonian is defined (up to rescaling and constant shift) as
\begin{equation}\label{eq:Hmc}
    H_{\text{MC}}(\graph) = \sum_{\langle i,j \rangle \in \mathcal{E}(\graph)} Z_i Z_j\, ,
\end{equation}
that is, similar to~\cref{eq:Htfim} for $h_X=0$, but, with interactions now depending on a potentially unstructured $\graph$.

The ground state of \cref{eq:Hmc} can be interpreted as the Max-Cut of $\graph$ (i.e., the bipartition of the set $\mathcal{V}(\graph)$ that maximizes the number of edges connecting each part).
As such, Max-Cut problems can be recasted as tasks of ground-state preparations, that are typically performed using QAOA circuits.
Similar to the HVA for the TFIM model, a layer of QAOA is composed of collective $ZZ$ rotations (now defined in correspondence to the set of edges $\mathcal{E}(\graph)$) and collective $X$ rotations, accounting for $n_l=2$ parameters per layer. 

\subsubsection{ Vanishing symmetries in random graphs }
Of importance to understand the following results, we recall that random graphs become almost surely asymmetric as their sizes increase. 
In other words, for large enough system sizes, a typical random graph admits no non-trivial automorphisms (see \cite{babai1995automorphism} and references thereof, for a more precise statement and assumptions).
In the context of constructing \orbitcircuit s for problems of Max-Cut, this implies that each node (edge) become structurally independent of any other node (edge), such that independent parameters are assigned to any of the parameterized gates composing the circuits. 
Hence, as the size of the underlying graphs are increased \orbitcircuit s would become more and more similar to Free circuits.

This aspect is illustrated in \cref{fig:mc_graphs} for $3$--regular graphs spanning system sizes $n=4$ to $20$ nodes.  
For $n=4$ nodes, the only $3$-regular graph is a complete graph, and due to the many symmetries present all the qubits (pairs of qubits) group into a single orbit (edge-orbit).
In this case the QAOA and \orbitcircuit\ are the same. When the size of the random graphs is increased, the number of symmetries quickly decrease, and the qubits (pairs of qubits) group into increasingly more distinct orbits (edge-orbits).
Already, for $n=10$ ($16$) nodes, this yields a total of $n_l=13$ ($37$) parameters.
This number still remains smaller than the number of parameters for the Free ansatz which is given by $n_l=|\mathcal{V}(\graph)|+|\mathcal{E}(\graph)|$ (i.e., a single parameter per node and per edge of $\graph$), and which for $3$-regular graphs equals $n+3n/2=25$ ($40$).
Finally, the graph with $n=20$ nodes only admits the identity as a trivial symmetry, such that each gate in \orbitcircuit\ has its own parameter (i.e., in this case ORB and Free circuits become effectively the same), in stark contrast with the $n_l=2$ parameters per layer of QAOA.
Overall, we see that ORB can recover both QAOA or Free circuits depending on the amount of symmetries present in the underlying graph, but in general will be different.

\subsubsection{ Results }
The performances of the three circuit constructions are reported in \cref{fig:mc_perf} for $\varepsilon=10^{-3}$.
As seen in~\cref{fig:mc_perf} (top panel), the critical number of layers for QAOA circuits is found to grow exponentially with $n$, and for $n>10$ nodes we could not achieve convergence given a maximum number of $1000$ layers. 
Still, both the ORB and Free circuits consistently converge within $10$ layers, with a small difference between their performances for $n \leq 10$ nodes. Such behavior is to be expected in light of the previous comments. 
As the size of the random graphs increases, performances of the Free and \orbitcircuit s become increasingly similar.
Remarkably, when looking at the critical number of parameters required (mid panel) and magnitude of the initial gradients (bottom panel), it can be seen that \orbitcircuit s systematically overperform other circuit constructions, gracefully interpolating between the best features of QAOA and Free circuits. 

Overall, the performance and trainability of \orbitcircuit s are found to be substantially better than those entailed by alternative circuit constructions. Again, we stress that, for the cases we studied, gradients are found to decrease exponentially fast with the system size, but, at a slower pace for \orbitcircuit . 

\subsubsection{ Discussion }\label{sec:maxcut_discusssion}
\begin{figure}[t!]
	\includegraphics[width=0.48\textwidth]{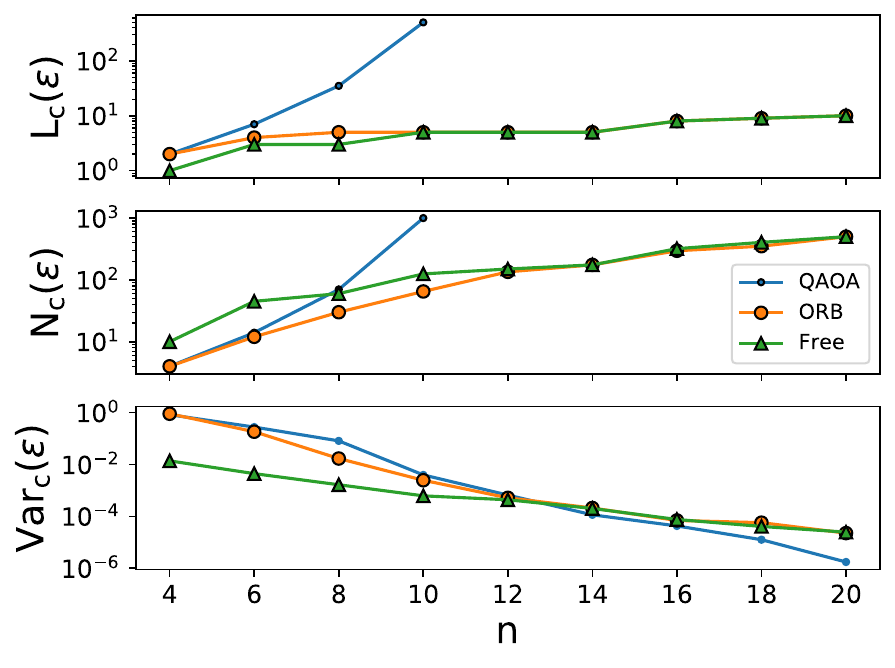}
	\caption{
    \textbf{Ground state preparation for Max-cut problems.} 
	The Max-Cut Hamiltonians in \cref{eq:Hmc} are defined in correspondence to randomly generated $3$-regular $n$-node graphs.	For each of the circuit constructions (colors in legend), we report the critical number of layers (top row), number of parameters (mid row), and gradient variance (bottom row) for $\varepsilon=10^{-3}$. 
	}
	\label{fig:mc_perf}
\end{figure}

Before proceeding to the final application, let us discuss ORB circuits in light of recent generalizations of the QAOA ones. 
In particular, the \emph{multi-angle} QAOA (ma-QAOA)~\cite{herrman2022multi} was recently presented and shown to overperform traditional QAOA for circuits with $L=1$ layers of ma-QAOA compared to up to $L=3$ layers for QAOA. 
In the context of QAOA, the Free circuits used for our benchmarks in~\cref{fig:mc_perf} are \emph{exactly} the ma-QAOA circuits of Ref.~\cite{herrman2022multi}. 
We now show that ORB circuits inherit the theoretical guarantees of ma-QAOA, and also improves on ma-QAOA as long as the underlying graphs under study retain some symmetries.

First, it is proven in Ref.~\cite{herrman2022multi} (Theorem 2.1 and Methods) that for $L\rightarrow \infty$ layers, there exist parameters of the ma-QAOA that exactly prepare ground states of the Max-Cut Hamiltonians in~\cref{eq:Hmc}. 
This follows directly from the fact that (i) in the large layers limit, the original QAOA circuits allow for such exact preparation~\cite{farhi2014quantum}, and that (ii) any QAOA circuit with $L$ layers can be exactly replicated by a ma-QAOA circuit with $L$ layers (simply by fixing each of the independent parameters of the ma-QAOA circuit to match the parameters of the QAOA one).  A similar argument readily extends to ORB-circuits showing that they also share the same guarantees for exact ground state preparations.

Second, it is proven in Ref.~\cite{herrman2022multi} (Theorem 4.1 and Methods) the systematic superiority of ma-QAOA over QAOA for arbitrary \emph{star graphs}. 
Precisely, it is shown that QAOA with $L=1$ layers ($2$ parameters) can never achieve perfect ground state preparation and have a performance ratio  $1-\varepsilon\rightarrow 75\%$ as the graph size $n$ increases.  
In contrast, one layer of ma-QAOA (containing $2n-1$ parameters) can exactly solve the problem for arbitrary $n$.

Recall that a $n$-node star graph is a graph consisting in $n-1$ outer nodes, each only connected to the remaining central node. 
Given the permutation symmetries of such graphs -- each of the $n-1$ outer nodes can be interchanged and thus belong to the same orbit --  \orbitcircuit s will require only $3$ parameters per layer, independently of the graph's size.
We verify that a single layer of \orbitcircuit s can also solve perfectly the problem. In fact, this already can be seen from the optimal parameters of the ma-QAOA provided in Ref.~\cite{herrman2022multi} (see Methods).
That is, we can certify that with almost $n$ times less parameters than ma-QAOA, and the same circuit depth, \orbitcircuit s achieve the same performances. 
This further shows the advantage of considering symmetries in the context of graph problems.

\subsection{$J_1$-$J_2$ Heisenberg model}\label{sec:J1J2}

\begin{figure*}
	\includegraphics[width=0.95\textwidth]{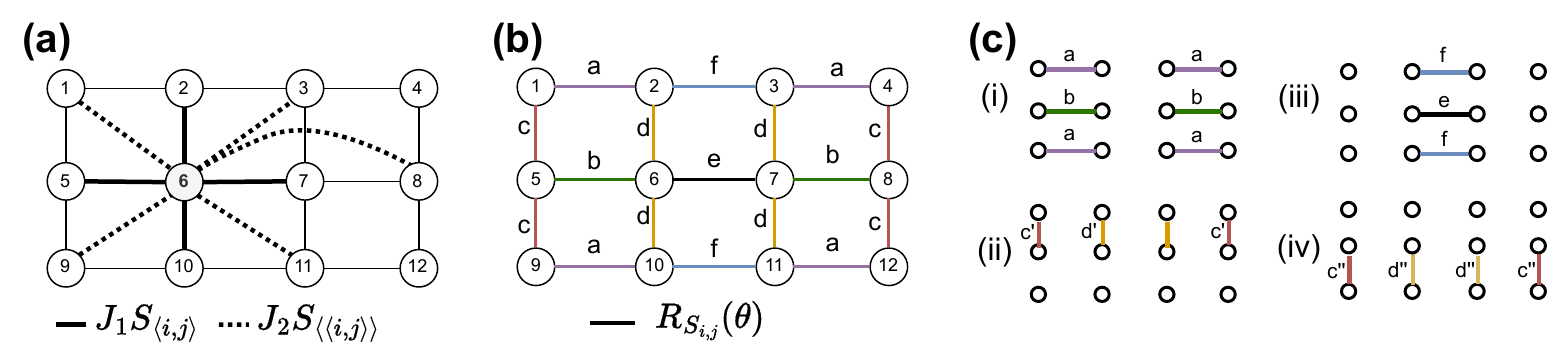}
	\caption{
    \textbf{$J_1$-$J_2$ Heisenberg model.}
    (a) The Hamiltonian, in~\cref{eq:J1J2}, is studied for a system arranged over a $3\times 4$ grid, and  is comprised of interactions $S_{i,j}$ between neighbors and next-neighbors with strengths $J_1$ and $J_2$ respectively (depicted here only for the $6$th node). 
    (b) For realistic implementation, we restrict the circuits to be composed of gates $R_{S_{i,j}}$ acting on pairs of neighbor qubits. 
    Given the grid configuration, these pairs separate as $6$ different edge-orbits (depicted in different colors and labelled as letters). 
    (c) A layer of \orbitcircuit\ is decomposed as $4$ sublayers (i-iv) of gates acting on non-overlapping pairs of qubits, and thus that can be realized in parallel.
    Given the non-commuting nature of the gates, the edge-orbits c and d, which contains intersecting edges, are further split yielding a total of $n_l=8$ parameters per layer.  
	}
	\label{fig:j1j2}
\end{figure*}

Going beyond models of matter involving only local interactions or commuting terms, \orbitcircuit\ readily extends to more challenging physical problems.
To showcase these possibilities, and address the practical issues arising in such cases, we conclude this section by studying the ground state preparation of a frustrated $J_1$-$J_2$ Heisenberg model~\cite{dagotto1989phase} laid on a grid.

The Hamiltonian of the system, for a $3\times4$ grid depicted in~\cref{fig:j1j2}(a), is defined as 
\begin{equation}\label{eq:J1J2}
    \ham_{J_1\text{-}J_2} = J_1 \sum_{\langle i,j \rangle} S_{ij} + J_2 \sum_{\langle \langle i,j \rangle \rangle} S_{ij},
\end{equation}
in terms of the Heisenberg exchange interactions $S_{ij}= X_iX_j + Y_iY_j + Z_iZ_j$ between pairs $\langle i,j \rangle$ of nearest-neighbor spins and pairs $\langle \langle i,j \rangle \rangle$ of next--nearest--neighbors, with strength $J_1$ and $J_2$ respectively. Due to these competing interactions, this model reveals a rich physics and remains subject of speculation w.r.t. the existence of a spin liquid phase, and potentially a second phase, in the vicinity of $J_2/J_1=0.5$~\cite{liu2018gapless}.
As such, it has been used as a testbed for the study of state-of-the-art numerical methods and the development of novel circuit constructions~\cite{bukov2021learning,zheng2021speeding,huerga2022variational,astrakhantsev2022algorithmic}. For example, in \cite{zheng2021speeding} an ansatz was proposed respecting the $\mathbb{SU}(2)$ invariance of~\cref{eq:J1J2} for the purpose of ground state preparation that will serve us as a benchmark.

In the following, we assume access to a quantum device with 2D neighbors-coupling (i.e., allowing for 2-qubit interactions between neighboring grid qubits). Given the $\mathbb{SU}(2)$ symmetry of~\cref{eq:J1J2}, we employ gates of the form $R_{S_{i,j}}=\exp [-i \theta S_{i,j}]$, over pairs of neighbors $\langle i,j \rangle$, that commute with the action of the symmetry group~\cite{zheng2021speeding,larocca2022group}. We stress that the set of gates employed here is only a subset of the terms appearing in~\cref{eq:J1J2}, but we do not require extended next-nearest-neighbor connectivity. 

The set of edge-orbits, restricted to pairs of neighbor qubits, is readily obtained by considering the two spatial symmetries of the grid (a horizontal and a vertical reflection).
This yields a total of $6$ edge-orbits labeled in~\cref{fig:j1j2}(b) from a to f.
While some of the edge-orbits only involve non intersecting edges (a, b, e, and f), this is not always the case (c and d). 
In practical terms, this means that due to the non-commuting nature of gates $R_{S_{i,j}}$ $R_{S_{i,k}}$ sharing a common qubit $i$, care needs to be taken when implementing ORB layers of the form~\cref{eq:genorb} under the assumption of $2$-qubit connectivity.

Consider as an example the two edges (1,5) and (5,9) belonging to the orbit c. \Cref{eq:genorb} would require to realize the unitary $R(\theta)=\exp[-i\theta (S_{1,5} + S_{5,9})]$,
which cannot be directly implemented as a composition of $R_{S_{1,5}}(\theta)$ and $R_{S_{5,9}}(\theta)$. 
Still, two options are possible. (i) One can compile $R(\theta)$ as gates $R_{S_{1,5}}$ and $R_{S_{5,9}}$ with suitable angles. 
Given that the Lie algebra generated by the terms $S_{1,5}$ and $S_{5,9}$ has dimension $4$, compilation of $R(\theta)$ would require a total of at least $4$ gates. 
(ii) As an alternative which does not increase the physical depth of a layer, one can slightly relax the equivariance requirement.
This is achieved by decorrelating parameters associated to overlapping gates belonging to a same edge-orbit, and results in our case in $2$ additional parameters, one for each of the edge-orbit c and d. 
Given our aim for shallow circuit constructions, we follow the second option.
Overall, the \orbitcircuit\ involves a total of $4$ rounds of gates that can be performed in parallel (we call such a round a sublayer), with a total of $n_l=8$ parameters per layer, depicted in~\cref{fig:j1j2}(c).

In comparison, a layer of HVA for such problem (extending the construction for the 1D Heisenberg chain in Ref.~\cite{ho2019efficient}) would also necessitate $4$ sublayers - this is the minimum number in order to realize a full set of $R_{S}$ gates over all the edges of the grid - and $n_l=4$ parameters. Such ansatz corresponds to the sublayers displayed in~\cref{fig:j1j2}(c) but with now a single parameter per sublayer (i-iv). The Free ansatz consists in the same sublayers with $1$ parameter per gate and a total of $n_l=17$ parameters per layer.

As studied in Refs.~\cite{ho2019efficient,wiersema2020exploring,zheng2021speeding}, the circuits are initialized as a tensor product of singlet states $\ket{s}=(\ket{01}-\ket{10})/\sqrt{2}$, which are defined here over adjacent qubits composing each row of the underlying grid.
Such a state has zero total spin and magnetization values that remain conserved during its evolution through the quantum circuit due to the choice of the $R_S$ gate-set. Furthermore, it can be verified that this initial state is invariant under the spatial symmetries of the grid.

For the frustrated case with $J_1=1$ and $J_2=0.5$ studied in~\cite{zheng2021speeding}, we achieve a relative ground state error $\varepsilon<5\times10^{-3}$ using \orbitcircuit\ with as few as $L=5$ layers, $85$ gates and $40$ parameters, a significant improvement compared to the circuit presented in Ref.~\cite{zheng2021speeding} that requires several hundreds of gates, most of them involving non-neighbor qubits (i.e., that would necessitate further extensive decomposition), and as many parameters.  Furthermore, and consistent with previous results, ORB circuit is found to perform better than HVA, which requires $L_c(\varepsilon)=30$ layers. It also performs well compared to the Free ansatz, which requires a similar number $L_c(\varepsilon)=4$ of layers, but twice as many parameters, and exhibits $5$ times smaller initial gradients.

\section{Summary and Outlook}\label{sec:conclusion}

In this work, we presented techniques for PQC construction based on careful considerations of the permutation symmetries of the underlying problem. Using tools from graph theory, we devised a principled way to correlate the parameters of the circuit, while maximizing the number of free parameters per layer and still ensuring invariance (under the relevant symmetries) of the realized states. We numerically probed the benefits of our construction, called \orbitcircuit\ (or ORB ansatz), in several tasks of ground state preparations. We  showed consistent improvement compared to HVA-like ansatzes and also compared to Free ansatzes which were designed for comparison purposes. While not solving issues of trainability at arbitrary system sizes, we expect that the proposed methods will help push further the range of problems that can be addressed with near-term wantum devices. 

As an unexpected result, we found that in many problems of combinatorial optimization \orbitcircuit\ construction advocates the use of independent parameters for each of the gates involved, contrasting with current practices of QAOA~\cite{farhi2014quantum}. 
The advantage of such approach was confirmed in~\cref{sec:mc} by means of numerical optimizations for problems of Max-Cut with random $3$-regular graphs of up to $20$ qubits, and study of trainability.

The construction presented here focuses on spatial symmetries, and as such, can be either used in complement with or can be decoupled from the internal symmetries of the problem. For instance, in~\cref{sec:J1J2}, invariance of the prepared state under the group of symmetries $\mathbb{SU}(2)$ was enforced by means of an adequate choice of the gate-set composing the anstaz.
Still, it has been remarked that breaking some of the internal symmetries of an ansatz could be beneficial in certain circumstances~\cite{bravyi2020obstacles,park2021efficient,choquette2020quantum}. 
In such cases, the principles presented could equally be applied to (internal) symmetry--breaking ansatzes~\cite{park2021efficient,choquette2020quantum}.
Similarly, \orbitcircuit s are equally portable to specialized ansatzes for problems of graph optimization extending the QAOA~\cite{hadfield2019quantum,cook2020quantum,golden2022evidence}.

While the focus of our work is on VQAs and Hamiltonian ground-state problems, our results are also applicable to the field of Quantum Machine Learning (QML)~\cite{schuld2015introduction,biamonte2017quantum} whereby one considers initial states stemming from a dataset. This will be particularly relevant for data with an underlying graph structure, as is the case in condensed matter and problems of graph classifications~\cite{verdon2019quantumgraph}. Moreover, our ORB ansatz will likely be useful in the context of VQAs for dynamical simulation~\cite{cirstoiu2020variational,commeau2020variational,gibbs2021long,gibbs2022dynamical}. In both QML and dynamical simulation, spatial symmetries appear and can be naturally accounted for with our ORB ansatz. We will explore these applications more in future work.
Finally, we conclude by discussing the relation of our work to prior work.

\subsubsection{Related works}

As was discussed in~\cref{sec:mc}, in the context of Max-Cut problems when the underlying graphs do not have any symmetries (as would most typically happen for large random graph) \orbitcircuit s naturally recover the ma-QAOA circuits~\cite{herrman2022multi}. 
However, when a few symmetries persist, ORB-circuits differ from ma-QAOA and we saw numerically its advantage. 
Furthermore, studies performed in this work were conducted in a different regime, with relatively large number of layers exceeding the numbers $L=1-3$ considered in Ref~\cite{herrman2022multi}.

Refs.~\cite{seki2020symmetry,astrakhantsev2022algorithmic} restore symmetries of a state prepared by a quantum circuit, which does not preserve symmetries of the problem, by means of additional measurements and post-processing, as opposed to the direct circuit construction presented here.

In Refs. \cite{shaydulin2021classical,shaydulin2021exploiting} the authors considered spatial symmetries and algebraic tools that are similar to the ones underpinning this work, but, limit their usage to performance predictions~\cite{shaydulin2021classical} or to the acceleration of numerical simulations of certain types of low--depth quantum circuits~\cite{shaydulin2021exploiting}. 

Finally, Ref.~\cite{meyer2022exploiting} proposes designing equivariant circuits through a process of gate symmetrization. We highlight that this symmetrization strategy can, when specialized to spatial symmetries, recover the construction proposed. In Ref.~\cite{meyer2022exploiting}, the merit of such symmetrization is assessed in the context of QML tasks, and also for fixed system-size problems of ground state preparations. Here, the benefit of incorporating spatial symmetries is demonstrated as a scaling effect, and its use is further theorized from an circuit expressibility perspective.

\section{Acknowledgements}

We thank Guillaume Verdon for insightful discussions and helpful feedback.  FS was supported by the Laboratory Directed Research and Development (LDRD) program of Los Alamos National Laboratory (LANL) under project number 20220745ER. ML and PJC were supported by the U.S. Department of Energy (DOE), Office of Science, Office of Advanced Scientific Computing Research, under the Accelerated Research in Quantum Computing (ARQC) program. ML was also supported by the Center for Nonlinear Studies at LANL.  PJC and MC acknowledge initial support from the LANL ASC Beyond Moore's Law project. MC was also supported by the Quantum Science Center (QSC), a National Quantum Information Science Research Center of the U.S. Department of Energy (DOE). 

\end{document}